\documentclass[11pt,a4paper]{article}
\pdfoutput=1
\usepackage{jheppub}
\usepackage[utf8]{inputenc}

\usepackage{color}
\usepackage{soul}
\usepackage{amsmath}
\usepackage{mathtools}
\usepackage{esint}
\usepackage{pifont}
\usepackage{bbold}

\usepackage{bbm}
\usepackage{verbatim}   
\usepackage{subfigure}
\usepackage{caption}
\usepackage{acronym}
\usepackage[export]{adjustbox}

\usepackage{amsfonts}
\usepackage{amssymb}
\usepackage{mathrsfs}
\usepackage{graphicx}
\usepackage{multirow}
\usepackage{slashed}

\usepackage{caption}

\definecolor{mypurple}{RGB}{164,64,214}
\title{The UV Fate of Anomalous $U(1)$s and the Swampland}

\author[a]{Nathaniel Craig,}
\emailAdd{ncraig@physics.ucsb.edu}
\author[b]{Isabel Garcia Garcia,}
\emailAdd{isabel@kitp.ucsb.edu}
\author[c]{and Graham D. Kribs}
\emailAdd{kribs@uoregon.edu}

\affiliation[a]{Department of Physics, University of California, Santa Barbara, CA 93106, USA}
\affiliation[b]{Kavli Institute for Theoretical Physics, University of California, Santa Barbara, CA 93106, USA}
\affiliation[c]{Department of Physics, University of Oregon, Eugene, OR 97403, USA}

\abstract{
Massive $U(1)$ gauge theories featuring parametrically light vectors are suspected to belong in the Swampland of consistent EFTs that cannot be embedded into a theory of quantum gravity.
We study four-dimensional, chiral $U(1)$ gauge theories that appear anomalous over a range of energies up to the scale of anomaly-cancelling massive chiral fermions.
We show that such theories require to be UV-completed at a finite cutoff below which a radial mode must appear, and cannot be decoupled --- a St\"uckelberg limit does not exist. When the infrared fermion spectrum contains a mixed $U(1)$-gravitational anomaly, this class of theories provides a toy model of a boundary into the Swampland, for sufficiently small values of the vector mass. In this context, we show that the limit of a parametrically light vector comes at the cost of a quantum gravity scale that lies parametrically below $M_{Pl}$, and our result provides field theoretic evidence for the existence of a Swampland of EFTs that is disconnected from the subset of theories compatible with a gravitational UV-completion. Moreover, when the low energy theory also contains a $U(1)^3$ anomaly, the Weak Gravity Conjecture scale makes an appearance in the form of a quantum gravity cutoff for values of the gauge coupling above a certain critical size.
}

\begin{document}

\maketitle

%%%%%%%%%%%%%%%%%%%%%%%%%%%%%%%%%%%%%%%%%%%%%%%%%%
%%%%%%%%%%%%%%%%%%%%%%%%%%%%%%%%%%%%%%%%%%%%%%%%%%
\section{Introduction}
\label{sec:intro}

Within the realm of effective field theory (EFT), certain features of continuous Abelian gauge theories starkly differ from those of their non-Abelian counterparts. Charge quantization is not `built-in', and the gauge group may be taken to be $\mathbb{R}$ as much as $U(1)$. Even if ratios of charges are assumed to be integer, arbitrarily large values appear consistent without the need to introduce an equally large number of degrees of freedom. However, some of these features are not expected to survive further UV-completion. Several arguments suggest that in a theory of quantum gravity Abelian charges must be quantized, and the corresponding gauge group compact \cite{Banks:2010zn}. To the extent that theories featuring large integer charge ratios approximately realize the non-compact limit --- abiding by the letter of the law but violating its spirit --- a shadow of suspicion hangs over those constructions.
In the language of our times, continuous Abelian gauge theories that exhibit some of these exotic features are expected to belong in the Swampland of consistent EFTs that cannot be UV-completed into a theory of quantum gravity \cite{Vafa:2005ui,ArkaniHamed:2006dz,Ooguri:2006in,Brennan:2017rbf}.

A further distinction between Abelian and non-Abelian gauge theories arises when the corresponding vector bosons are massive. In massive Yang-Mills, the breakdown of perturbation theory, manifest in the loss of perturbative unitarity in longitudinal gauge boson scattering, requires that the theory be UV-completed at scales of order $\Lambda \lesssim 4 \pi m / g$ (up to group theoretic factors). On the other hand, a massive Abelian gauge theory coupled to a conserved current is renormalizable: the photon mass $m_\gamma$ and the gauge coupling strength $g$ are free parameters of the theory --- ignoring Landau poles, such a theory may be valid up to arbitrarily high scales \cite{Proca:1900nv,Stueckelberg:1938zz,Boulware:1970zc,Salam:1971sp}.

It is easy to see that an Abelian gauge theory coupled to a conserved current is renormalizable even when the photon is massive. We may start with an Abelian Higgs model, where the photon mass arises as a result of spontaneous symmetry breaking. In the broken phase, a scalar excitation -- the Higgs `radial mode' -- is part of the spectrum, with mass proportional to the vacuum expectation value (vev) of the Higgs field. The low energy theory, featuring a massive photon coupled to charged matter, will therefore break down at the scale at which this radial mode appears. However, we may fully decouple the radial mode by taking the limit of infinite Higgs vev, $f \rightarrow \infty$. Doing so while keeping the photon mass finite requires that the charge of the Higgs field must be simultaneously taken to zero. If charged matter is to remain coupled to the photon is this limit, then the ratio of charges between the infrared charged spectrum and the Higgs must diverge.
%In other words: this limit is formally possible, but only if we are willing to let the gauge group be $\mathbb{R}$ and not $U(1)$.
This implementation is often referred to as the `St\"uckelberg limit' of a massive Abelian gauge theory. Even if we moderate our ambitions, and only allow the scale of spontaneous symmetry breaking to grow as high as $f \sim M_{Pl}$, we might still make the photon lie as far below the scale of the radial mode as we want by taking the Higgs charge tiny. A large, albeit potentially integer, hierarchy of charges ensures that charged matter remains coupled to the photon in this limit, while allowing for the gauge group to remain compact.

The above discussion makes it clear that there cannot be a model-independent upper bound on the cutoff scale of a massive Abelian gauge theory coupled to a conserved current. However, it also highlights how theories with massive photons that feature parametrically high cutoffs are highly suspect: decoupling additional degrees of freedom related to the dynamical mechanism that generates a photon mass requires introducing a parametrically large ratio of charges that we suspect is not allowed within a gravitational UV-completion.

In light of the above, one is prompted to ask: Do tiny photon masses belong in the Swampland? This question was the focus of \cite{Reece:2018zvv}, where a first attempt was made to understand the difficulties of realizing parametrically small photon masses in UV-completions that include gravity by studying the properties of string theory constructions where the photon mass is non-zero everywhere in field space, and which appear qualitatively different from the Abelian Higgs model.
\footnote{Implementations of a massive Abelian gauge theory such that at no point in field space the photon mass vanishes are commonly referred to in the string literature as `St\"uckelberg masses', which is the terminology employed in \cite{Reece:2018zvv}. By contrast, in this paper we will simply use the term `St\"uckelberg limit' to refer to the limit where a radial mode related to the mechanism that generates the photon mass can be fully decoupled.}
Heuristic arguments involving the expectation that a large number of degrees of freedom must become part of the low energy theory when wandering over large distances in field space \cite{Ooguri:2006in}, as well as the corresponding lowering of the quantum gravity cutoff in a theory with a large number of species \cite{Dvali:2007hz,Dvali:2007wp}, make for a compelling argument that the limit of tiny photon masses is problematic. %Drawing from various conjectures established to varying levels of rigour, \cite{Reece:2018zvv} further suggests that such constructions must feature a non-decoupling radial mode, and advocates for demanding that the theory satisfies the Weak Gravity Conjecture (WGC) \cite{ArkaniHamed:2006dz} even though the photon mass is non-vanishing.
\cite{Reece:2018zvv} further suggests that such constructions must feature a non-decoupling radial mode, and advocates for demanding that the theory satisfies the Weak Gravity Conjecture (WGC) \cite{ArkaniHamed:2006dz} even though the photon mass is non-vanishing.

Here, we will be concerned with an \emph{a priori} unrelated class of theories for which the limit $m_\gamma \rightarrow 0$ is also singular: four-dimensional, chiral gauge theories whose infrared fermion content is anomalous. Apart from being interesting in their own right, this class of theories frequently arise in the low energy limit of string theory constructions \cite{Dine:1987xk,Aldazabal:1998mr,Ibanez:1998qp,Ibanez:1999pw,Antoniadis:2002cs}. An anomalous gauge theory can be consistently quantized in perturbation theory so long as gauge bosons are massive, and the theory is further UV-completed at some finite cutoff scale \cite{Krasnikov:1985bn,Faddeev:1986pc,Preskill:1990fr}. For a non-Abelian theory the role of the anomaly is incidental: it forces the gauge bosons to acquire a mass, but plays no role in the cutoff size --- instead, the non-renormalizability of massive Yang-Mills already sets an upper bound $\Lambda \lesssim 4 \pi m / g$. On the other hand, the upper bound on the cutoff of an anomalous Abelian gauge theory is set purely by the anomaly. In keeping with the renormalizability of the Abelian non-linear sigma model, as the effects of the anomaly vanish, any upper bound on the cutoff of the anomalous EFT correspondingly disappears. In four dimensions, the cutoff of the anomalous EFT corresponds to the scale at which massive fermions appear, with the appropriate charge assignments to cancel the anomalies of the low energy spectrum.

This disparity between Abelian and non-Abelian gauge theories becomes particularly significant in the presence of gravity. Contrary to their non-Abelian counterparts, chiral Abelian gauge theories may present a mixed gravitational anomaly in four dimensions. In such case, the anomaly implies an upper bound on the cutoff scale of the anomalous EFT, of the form $\Lambda_{\rm grav} \sim 4 \pi \left( {M}_{Pl}^2 m_\gamma / | {\rm tr} (g_i) | \right)^{1/3}$ \cite{AlvarezGaume:1983ig,Preskill:1990fr}. The effect decouples in the limit $M_{Pl} \rightarrow \infty$ where gravity is turned off, whereas $\Lambda_{\rm grav} \rightarrow 0$ in the limiting of vanishing photon mass --- as advertised, the limit $m_\gamma \rightarrow 0$ is not allowed. For finite $M_{Pl}$, this class of theories provide a rare, field-theoretic toy model of a boundary into the Swampland, by setting an upper bound on the scale of additional degrees of freedom required for theoretical consistency in the presence of gravity.

\begin{figure}
  \centering
  \includegraphics[scale=1.2]{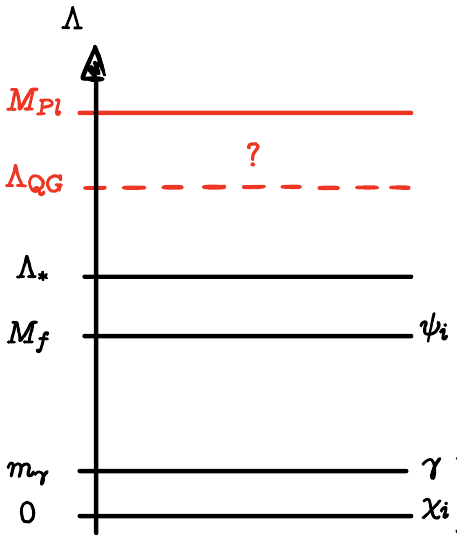}
  \captionof{figure}{Illustration of the different scales relevant to our discussion. At low energies, the anomalous EFT contains a number of massless fermions ($\chi_i$), and a massive photon. Massive fermions ($\psi_i$) responsible for cancelling the anomalies of the low energy spectrum appear at scale $M_f$. This anomaly-free extension may require further UV-completion at some higher scale $\Lambda_*$. In the presence of gravity, the quantum gravity scale, which may be below $M_{Pl}$, will play an interesting role in our discussion.}
  \label{fig:scales}
\end{figure}

Motivated by this unique feature of Abelian gauge theories, in this paper we focus on chiral $U(1)$ gauge theories that are anomaly-free, but for which anomaly cancellation occurs due to fermions appearing at different scales. As illustrated in Figure \ref{fig:scales}, these theories, which represent partial UV completions of the anomalous EFTs that are the focus of \cite{Preskill:1990fr}, feature a variety of mass scales above the photon mass. When gravitational effects are decoupled, the most relevant scales are the masses $M_f$ of the heavy fermions responsible for anomaly-cancellation, as well as a possible cutoff $\Lambda_*$ of the anomaly-free theory. When gravitational effects are included, the four-dimensional Planck scale, and the quantum gravity scale $\Lambda_{\rm QG}$ (which may differ from $M_{Pl}$) also enter into the discussion.
 %%%%%%%%%%%%%%%%%%%%
%%%%%%%%%%%%%%%%%%%%
\emph{The main focus of this paper is to explore the properties of this class of chiral gauge theories, paying special attention to the consequences of probing the regime of a parametrically light vector.} In doing so, we show that:
\begin{itemize}
	\item[(\emph{i})] This class of massive Abelian gauge theories are themselves EFTs, and require further UV-completion at a finite scale $\Lambda_*$ above the scale of the heavy fermions. This cutoff scale corresponds to an upper bound on the mass of a radial mode that cannot be decoupled --- \emph{in this class of theories, a St\"uckelberg limit does not exist}.
\end{itemize}
The above result is a consequence of the presence of massive fermions with chiral charge assignments necessary to cancel the anomalies of the low energy EFT. The loss of perturbative unitarity in fermion--anti-fermion annihilation into a number of longitudinal photons sets an upper bound on the scale of UV-completion required to recover a perturbative expansion.
An additional degree of freedom must become part of the spectrum, and we will generically refer to it as a `radial mode', although we emphasize that it need not correspond to the radial mode of a weakly coupled Abelian Higgs UV-completion.
Although the limit $\Lambda_* \rightarrow \infty$ is not accessible, a parametric separation of scales $m_\gamma \ll g \Lambda_* / 4 \pi$ remains possible, but only at the cost of introducing a large hierarchy of charges within the UV-completion. The consequences of introducing a large hierarchy of charges as required to realize a parametrically light vector differ dramatically depending on whether the low energy theory features a mixed $U(1)$-gravitational anomaly. When this is the case, we find that:
\begin{itemize}

	\item[(\emph{ii})]
	Introducing a large hierarchy of charges necessarily requires the presence of an equally large number $N$ of massive fermion species to cancel the anomaly of the low energy spectrum. Correspondingly, the quantum gravity scale is lowered down to $\Lambda_{\rm QG} \sim 4 \pi {M}_{Pl} / \sqrt{N}$ \cite{Dvali:2007hz,Dvali:2007wp}. \emph{Thus, probing the (approximately) non-compact limit that is required to realise a parametrically small photon mass comes at the cost of a quantum gravity cutoff that lies parametrically below ${M}_{Pl}$. }
	
	\item[(\emph{iii})] Although the cutoff scale $\Lambda_*$ of the anomaly-free theory cannot be decoupled, it may be pushed all the way up to the quantum gravity scale. When this is possible, and the limit is saturated:
	\begin{equation} \label{eq:1}
		\Lambda_* \sim \Lambda_{\rm QG} \sim 4 \pi \left( \frac{{M}_{Pl}^2 m_\gamma}{g} \right)^{1/3} \ .
	\end{equation}
	The right-hand-side of Eq.(\ref{eq:1}) parametrically coincides with the scale $\Lambda_{\rm grav}$ below which the massive, anomaly-cancelling fermions must appear. To the extent that theories that remain anomalous above this scale would belong in the Swampland of consistent non-gravitational EFTs that cannot be coupled to gravity, our result shows that such a possibility is self-consistently avoided by the lowering of the quantum gravity scale as mandated by the presence of a large number of species.
\emph{This class of theories therefore provides field-theoretic evidence for the existence of a Swampland of EFTs that is disconnected from the subset of theories that are compatible with a gravitational UV-completion.}
\footnote{Of course, there are more consistency conditions an EFT needs to satisfy in order to be compatible with an underlying theory of quantum gravity than those stemming from the requirement that gravitational anomalies are cancelled, and it is the goal of the Swampland Program to identify and understand all of those conditions. However, whereas a consistent EFT would need to satisfy all of those requirements, it is enough to fail one to fall into the Swampland.}
	
	\item[(\emph{iv})] In the more generic case where the low energy fermion spectrum features both $U(1)^3$ and $U(1)$-gravitational anomalies, there exists a critical value of the gauge coupling, given by
	\begin{equation}
		g_* \sim \left( \frac{64 \pi^3 m_\gamma}{{M}_{Pl}} \right)^{1/4} \ .
	\label{eq:gstar}
	\end{equation}
	When $g \lesssim g_*$ the cutoff scale of the anomaly-free EFT may be taken as high as the quantum gravity cutoff, and the statements in (\emph{iii}) apply. On the other hand, when $g \gtrsim g_*$ the upper bound on $\Lambda_*$ lies parametrically below the quantum gravity scale. In this case, a large separation of scales $m_\gamma \ll g \Lambda_* / 4 \pi$ comes at the cost of lowering the quantum gravity cutoff down to
	\begin{equation}
		\Lambda_{\rm QG} \sim g {M}_{Pl} \ ,
	\end{equation}
	which parametrically coincides with the WGC scale as seen from the low energy EFT \cite{ArkaniHamed:2006dz}. Moreover, in this regime $g \sim g_0^{1/3}$, where $g_0$ is the charge quantum of the anomaly-free theory, and therefore $\Lambda_{\rm QG} \sim g_0^{1/3} {M}_{Pl}$, which is the version of the magnetic WGC scale advocated for in \cite{Heidenreich:2016aqi,Heidenreich:2017sim}. \emph{This provides a four-dimensional, field-theoretic example of a class of massive Abelian gauge theories where the WGC scale emerges in the role of a quantum gravity cutoff in a way that is tied to the presence of a large number of species.}
	
 \end{itemize}

This paper is organized as follows. In section~\ref{sec:review} we review some of the properties of anomalous Abelian gauge theories in four dimensions that are most relevant to our discussion, following \cite{Preskill:1990fr}. In \ref{sec:unitarity}, we show how the presence of massive fermions with chiral charge assignments leads to the breakdown of perturbation theory at high energies, calling for further UV-completion of the anomaly-free theory. We discuss how the St\"uckelberg limit in which the upper bound on the cutoff is decoupled is not accesible in this class of models. In section~\ref{sec:grav} we focus on the implications of our results for Abelian gauge theories featuring anomalous fermion content at low-energies, with special attention to the implication of mixed $U(1)$-gravitational anomalies. Section~\ref{sec:discussion} contains our conclusions.

%%%%%%%%%%%%%%%%%%%%%%%%%%%%%%%%%%%%%%%%%%%%%%%%%%
%%%%%%%%%%%%%%%%%%%%%%%%%%%%%%%%%%%%%%%%%%%%%%%%%%
\section{EFT cutoffs in anomalous Abelian gauge theories}
\label{sec:review}

Unlike massive Yang-Mills, a non-zero gauge boson mass in the context of an Abelian gauge theory does not lead to the breakdown of perturbation theory at high external momenta. Although not obvious in unitary gauge, where the gauge boson propapagator falls off slower than $1/k^2$, it becomes apparent if we enlarge the theory so as to introduce a gauge redundancy by incorporating an additional degree of freedom $\theta$ transforming non-linearly under the gauge action. This allows us to rewrite the vector mass term as
\begin{equation} \label{eq:massiveU1}
	\mathcal{L} \supset	\frac{1}{2} m_\gamma^2 A_\mu^2 \quad \rightarrow \quad \frac{1}{2} (\partial_\mu \theta - m_\gamma A_\mu)^2 \ ,
\end{equation}
which remains invariant under gauge transformations of the form $A_\mu \rightarrow A_\mu + \frac{1}{g_0} \partial_\mu \alpha$ and $\theta \rightarrow \theta + \frac{m_\gamma}{g_0} \alpha$, where $g_0$ refers to the unit of electric charge. This is the so-called `St\"uckelberg trick' --- its crucial insight being that it is possible to restore gauge invariance without introducing operators of dimension higher than 4, making the renormalizability of a massive Abelian gauge theory manifest \cite{Stueckelberg:1938zz}. This remains true if any fermionic current that $A_\mu$ couples to is vector-like, regardless of the fermion mass. In this case, $m_\gamma$ is a free parameter of the theory, and the limit $m_\gamma \rightarrow 0$ remains unproblematic.

This is no longer true if the current that $A_\mu$ couples to is not conserved, such as in the context of theories with anomalous fermion content. Nevertheless, for both Abelian and non-Abelian groups, gauge theories with anomalies can be consistently quantized, so long as the corresponding vector bosons are massive, and that the theory is treated as an EFT only valid up to a finite cutoff scale \cite{Krasnikov:1985bn,Faddeev:1986pc,Preskill:1990fr}. For an Abelian gauge theory, the upper bound on the EFT cutoff depends solely on the anomaly, and differs parametrically for $U(1)^3$ and mixed $U(1)$-gravitational anomalies. In the remainder of this section, we review the status of theories with Abelian gauge anomalies, closely following \cite{Preskill:1990fr}.

For illustration, we focus on a theory containing a single massless fermion, coupled to $A_\mu$ through a left-handed current. Allowing for a non-zero photon mass, the corresponding lagrangian reads
\begin{equation} \label{eq:anomalousL}
\mathcal{L} = - \frac{1}{4} F_{\mu \nu} F^{\mu \nu} + \frac{1}{2} (\partial_\mu \theta - m_\gamma A_\mu)^2 + \bar \chi i \gamma^\mu \partial_\mu \chi + g A_\mu \bar \chi_L \gamma^\mu \chi_L \ ,
\end{equation}
where $g \equiv g_0 Q$. \footnote{Distinguishing between $g_0$ and $g$ may seem unnecessary at this point. However, it will be relevant in our subsequent discussion, where we will consider extensions of the anomalous EFT featuring a quantum of charge that differs from the typical gauge coupling of the infrared spectrum.} Although Eq.(\ref{eq:anomalousL}) remains invariant under gauge transformations, under which the left- and right-handed components of $\chi$ transform as $\chi_L \rightarrow e^{ i Q \alpha} \chi_L$ and $\chi_R \rightarrow \chi_R$, the corresponding path integral does not due to the non-trivial jacobian of the fermionic functional determinant. Effectively, the presence of a $U(1)^3$ anomaly leads to an additional term in the lagrangian, of the form
\begin{equation} \label{eq:anomalousdL}
	\delta \mathcal{L} = \frac{1}{3} \frac{g_0^2 Q^3}{16 \pi^2} \alpha F \tilde F \ .
\end{equation}
At this point, one could try to restore gauge invariance by modifying the theory into an anomaly-free one, e.g.~by introducing a coupling between $A_\mu$ and the right-handed fermion current with identical strength, rendering the entire interaction vector-like. Alternatively, one could choose to leave the theory as it is, and instead build a gauge invariant version of the anomalous EFT by adding a term to the lagrangian proportional to $\theta F \tilde F$ with the appropriate coefficient to cancel Eq.(\ref{eq:anomalousdL}):
\begin{equation} \label{eq:Ltheta3}
	\mathcal{L} \supset - \frac{1}{3} \frac{g^3}{16 \pi^2 m_\gamma} \theta F \tilde F \ .
\end{equation}
This is the Abelian version of the Wess-Zumino term \cite{Wess:1971yu,Witten:1983tw} --- Eq.(\ref{eq:anomalousL}) extended with this new term provides a gauge invariant description of our anomalous EFT. However, it is apparent from Eq.(\ref{eq:Ltheta3}) that gauge invariance of the anomalous theory has only been achieved at the cost of renormalizability. Moreover, the coefficient of the $\theta F \tilde F$ term diverges in the limit $m_\gamma \rightarrow 0$, which provides an easy way to see that the limit of a massless photon is indeed not allowed in an anomalous theory. The scale suppressing this operator corresponds to the cutoff of the anomalous EFT. Following standard NDA counting \cite{Manohar}, an upper bound on the scale of UV-completion as mandated by the presence of a $U(1)^3$ anomaly is given by \cite{Preskill:1990fr}
\begin{equation} \label{eq:Lambda3}
	\Lambda_{U(1)^3} \sim \frac{64 \pi^3 m_\gamma}{g^3} \ .
\end{equation}

Turning on gravity, Eq.(\ref{eq:anomalousdL}) is accompanied by an extra term $\propto \alpha R \tilde R$ due to the presence of a mixed $U(1)$-gravitational anomaly.
%\begin{equation}
%	\delta \mathcal{L}_{\rm grav} = \frac{1}{24} \frac{Q}{16 \pi^2} \sqrt{|g|} \alpha R \tilde R \ .
%\end{equation}
As before, gauge invariance may be restored in perturbation theory by including the following term:
\begin{equation}
	\mathcal{L} \supset - \frac{1}{24} \frac{g}{16 \pi^2 m_\gamma} \sqrt{|{\rm det} g|} \theta R \tilde R \ .
\end{equation}
The scale suppressing this operator sets an upper bound on the scale of UV-completion as mandated by the presence of a mixed $U(1)$-gravitational anomaly. Parametrically \cite{AlvarezGaume:1983ig,Preskill:1990fr}:
\begin{equation} \label{eq:Lambdagrav}
	\Lambda_{\rm grav} \sim 4 \pi \left( \frac{{M}_{Pl}^2 m_\gamma}{g} \right)^{1/3} \ .
\end{equation}

\begin{figure}
  \centering
  \includegraphics[scale=1.2]{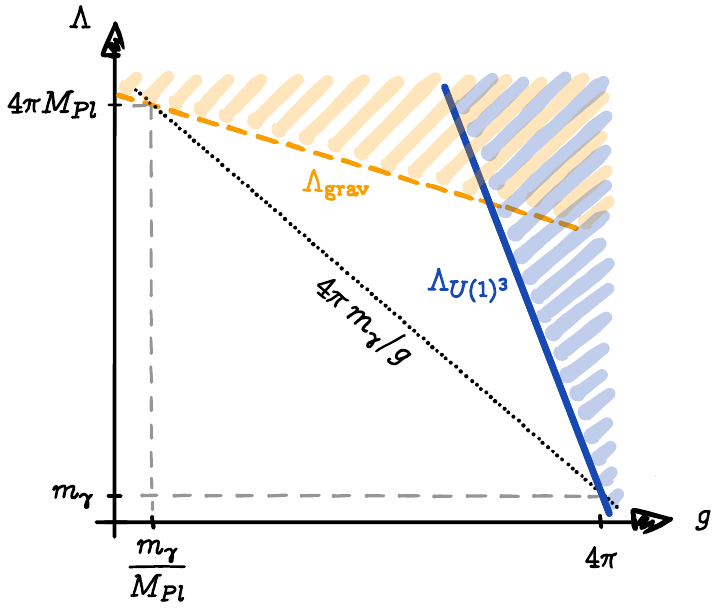}
  \caption{Upper bound on the cutoff of a massive Abelian gauge theory coupled only to a left-handed fermion current. The solid blue and dashed orange lines correspond to the scale at which perturbation theory breaks down as a result of the presence of a $U(1)^3$ and a mixed $U(1)$-gravitational anomaly respectively, as given in Eq.(\ref{eq:Lambda3}) and (\ref{eq:Lambdagrav}). The value of $g$ at which the two lines meet is given in Eq.(\ref{eq:gstar}). To the extent that theories living in the orange shaded region (above the orange dashed line but below the solid blue line) correspond to consistent non-gravitational EFTs that become inconsistent in the presence of gravity, the orange wedge represents a piece of Swampland. Notice that the upper bound on the EFT cutoff always lies above the scale $4 \pi m_\gamma / g$, further highlighting the qualitative difference between massive Abelian gauge theories and their non-Abelian counterparts. (Both axes are in a log scale.)}
  \label{fig:cutoff}
\end{figure}

In a four-dimensional theory, Eq.(\ref{eq:Lambda3}) and (\ref{eq:Lambdagrav}) correspond to the scale below which massive fermions must appear, with charge assignments appropriate to cancel the corresponding anomaly. As depicted in Figure~\ref{fig:cutoff}, the cutoff of the anomalous EFT may lie parametrically above the scale $4 \pi m_\gamma / g$ for all perturbative values of the gauge coupling, and so long as the implied cutoff falls below the gravitational scale (that is, $m_\gamma / {M}_{Pl} \lesssim g \lesssim 4 \pi$). The special value of the gauge coupling advertised in Eq.(\ref{eq:gstar}) already makes an appearance here: the upper bound on the cutoff scale of the anomalous EFT is dominated by either the mixed gravitational anomaly or the $U(1)^3$ anomaly depending on whether $g \lesssim g_*$ or $g \gtrsim g_*$, respectively. (We will have more to say about the behaviour of the quantum gravity scale in these two regimes in section \ref{sec:both}.). Eq.(\ref{eq:Lambda3}) and (\ref{eq:Lambdagrav}) can be adapted to a more general fermion content after the respective substitutions $g^3 \rightarrow g_0^3 \left| {\rm tr} ( Q_i^3 ) \right|$, and $g \rightarrow g_0 \left| {\rm tr} ( Q_i ) \right|$. However, in the absence of large hierarchies of charges in the low energy spectrum, Eq.(\ref{eq:Lambda3}) and (\ref{eq:Lambdagrav}) will still provide a parametrically correct estimate of the upper bound on the EFT cutoff, for theories featuring the corresponding anomaly, with $g$ being understood as the typical size of the gauge coupling present in the infrared.
 %%%%%%%%%%%%%%%%%%%%
%%%%%%%%%%%%%%%%%%%%

\begin{figure}
  \centering
  \includegraphics[scale=1.2]{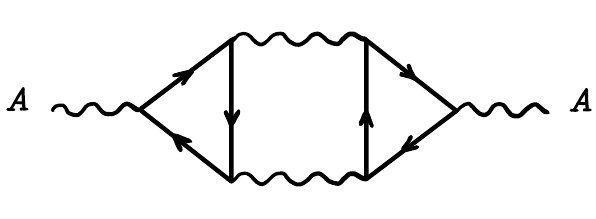}
  \caption{A non-zero radiative contribution to the photon mass, parametrically of the form Eq.(\ref{eq:massanomaly}), arises from this three-loop diagram, with massive fermions responsible for cancelling the $U(1)^3$ anomaly of the low energy theory propagating inside the loops \cite{Preskill:1990fr}. If the low energy theory also contains a mixed $U(1)$-gravitational anomaly, there is an analogous diagram with gravitons (instead of vectors) propagating in the internal lines.}
  \label{fig:triangle}
\end{figure}

Eq.(\ref{eq:Lambda3}) and (\ref{eq:Lambdagrav}) were first obtained in \cite{Preskill:1990fr} and \cite{AlvarezGaume:1983ig} respectively, and, as sketched above, can be derived within the anomalous EFT alone. However, they can be readily understood by considering the effect of heavy fermions with mass $M_f$ that must be present in any four-dimensional UV-completion in order to render the full theory anomaly-free. Through the diagram depicted in Figure~\ref{fig:triangle}, the heavy fermions responsible for cancelling the $U(1)^3$ anomaly lead to a non-zero contribution to the photon mass, of the form \cite{Gross:1972pv,Preskill:1990fr}
\begin{equation} \label{eq:massanomaly}
	\delta m_\gamma^2 \sim \left( \frac{g^3}{64 \pi^3} M_f \right)^2 \ .
\end{equation}
The requirement that $m_\gamma^2 \gtrsim \delta m_\gamma^2$ yields Eq.(\ref{eq:Lambda3}), with the role of $\Lambda_{U(1)^3}$ played by the mass of the heavy fermions. Crucially, as discussed in \cite{Preskill:1990fr}, this is more than a statement about the natural size of $m_\gamma$ --- fine-tuning the photon mass below $\delta m_\gamma$ would require fine-tuning the coefficients of an infinite number of higher-dimensional-operators, effectively signalling the breakdown of perturbation theory within the anomalous EFT. Identical considerations apply to a version of Figure \ref{fig:triangle} with gravitons propagating in the internal lines in theories with a mixed gravitational anomaly, and similarly lead to the scale in Eq.(\ref{eq:Lambdagrav}) being identified with an upper bound on the mass of the anomaly-cancelling fermions.
 %%%%%%%%%%%%%%%%%%%%
%%%%%%%%%%%%%%%%%%%%

%%%%%%%%%%%%%%%%%%%%%%%%%%%%%%%%%%%%%%%%%%%%%%%%%%
%%%%%%%%%%%%%%%%%%%%%%%%%%%%%%%%%%%%%%%%%%%%%%%%%%
\section{Chiral Abelian gauge theories and massive fermions}
\label{sec:unitarity}

We now turn our attention to the ultraviolet fate of the anomalous Abelian gauge theories discussed in section~\ref{sec:review}, cancelling the anomalies of the low energy theory by including massive fermions with appropriate quantum numbers, which is the only option in four dimensions. For simplicity, we focus first on a single heavy fermion $\psi$ with mass $M_f$ and chiral $U(1)$ charges. The terms in the lagrangian involving $\psi$ read
\begin{equation}
\mathcal{L} \supset \bar \psi i \gamma^\mu \partial_\mu \psi + g_L A_\mu J^\mu_L + g_R A_\mu J^\mu_R - M_f (\bar \psi_L \psi_R + {\rm h.c.}) \ ,
\end{equation}
where
\begin{equation} \label{eq:Jpsi}
	J^\mu_X \equiv \bar \psi_X \gamma^\mu \psi_X \quad {\rm and} \quad g_X \equiv g_0 Q_X \quad {\rm for} \quad X = L, R \ ,
\end{equation}
and $Q_L \neq Q_R$ in general. By assumption, $Q_L$ and $Q_R$ are such that the full theory is free of both $U(1)^3$ and mixed gravitational anomalies. Thus, when $Q_L = Q_R$, and the heavy fermion couples to $A_\mu$ through a vector current, the massless fermion sector must itself be non-anomalous. However, when $Q_L \neq Q_R$ the heavy fermion will play a role in anomaly cancellation, and the theory will appear anomalous below the scale $M_f$.

Having enlarged the fermion spectrum so as to make the theory anomaly-free, a gauge transformation leaves the fermionic functional determinant in the path integral unchanged. However, it is the fermion mass term that now breaks gauge invariance, which we may restore by introducing appropriate couplings to $\theta$, as follows:
\begin{equation} \label{Eq:fermionmass}
	\mathcal{L} 	%\supset	- M_f \left( e^{- \frac{i \theta}{f}} \bar \psi_R \psi_L + {\rm h.c.} \right)
				\supset	- M_f \left( e^{\pm \frac{i \theta}{f}} \bar \psi_L \psi_R + {\rm h.c.} \right)
				%= - M_f \left[ \cos \left( \frac{\theta}{f} \right) \bar \psi \psi + i \sin \left( \frac{\theta}{f} \right) \bar \psi \gamma^5 \psi \right] \ ,
				= - M_f \left[ \cos \left( \frac{\theta}{f} \right) \bar \psi \psi \pm i \sin \left( \frac{\theta}{f} \right) \bar \psi \gamma^5 \psi \right] \ ,
\end{equation}
where
\begin{equation} \label{eq:f}
%	f \equiv \frac{m_\gamma}{g_0 (Q_L - Q_R)} \ .
	f \equiv \frac{m_\gamma}{| g_L - g_R |} = \frac{m_\gamma}{g_0 | Q_L - Q_R |}\ ,
\end{equation}
and the upper (lower) sign in Eq.(\ref{Eq:fermionmass}) applies when $Q_L - Q_R >0$ ($Q_L - Q_R < 0$).
Expanding the sine and cosine in Eq.(\ref{Eq:fermionmass}) as an infinite sum of higer-dimensional-operators suppressed by increasing powers of $f$ suggests that our attempt to restore gauge invariance has only been successful at the cost of renormalizability, and one might expect the theory to be valid only up to scales not much above $f$ itself. Indeed, any attempt to rewrite Eq.(\ref{Eq:fermionmass}) in a way that involves only renormalizable interactions necessarily requires introducing additional degrees of freedom. For instance, if the theory is further embedded into a weakly coupled Abelian Higgs model with condensate charge $Q_L - Q_R$ and vev $f$, then a radial mode will be present below the scale $4 \pi f$. The UV-completion of the anomalous EFT by the addition of massive fermions is {\it itself} an effective field theory. In what follows, we will refer to this partial UV-completion as the {\it anomaly-free EFT}. We now confront a nested set of EFTs: the anomalous EFT valid up to the scale $M_f$, and the non-anomalous EFT valid from the scale $M_f$ up to some cutoff $\Lambda_*$. Several questions then arise: what is the cutoff of the anomaly-free EFT? When does it coincide with the apparent cutoff of the anomalous EFT? And when, if ever, can we take $\Lambda_* \rightarrow \infty$? These are the questions that we address in this section.

It is clear from the form of Eq.(\ref{Eq:fermionmass}) that there are at least {\it some} cases in which the limit $\Lambda_* \rightarrow \infty$ will be allowed and the theory becomes fully renormalizable, regardless of the value of the photon mass: (\emph{i}) when $M_f \rightarrow 0$, regardless of the left- and right-handed charge assignments, and (\emph{ii}) when $Q_R \rightarrow Q_L$, regardless of the fermion mass. In both cases, any upper bound on $\Lambda_*$ due to Eq.(\ref{Eq:fermionmass}) must decouple, and a St\"uckelberg limit must exist.

However, to the extent that Eq.(\ref{Eq:fermionmass}) involves irrelevant operators, one might wonder whether the description of the non-anomalous EFT should be enlarged to include additional irrelevant operators compatible with the gauge symmetry, whose appearance might lead to an independent bound on $\Lambda_*$, and additional conditions on the realization of a St\"uckelberg limit. The status of such operators can be readily ascertained from a chiral lagrangian analysis, which we carry out in section \ref{sec:EFT}. In \ref{sec:tree}, we show how, whenever $M_f \neq 0$ and $Q_L \neq Q_R$, the loss of perturbative unitarity at high energies signals the breakdown of perturbation theory, and leads to an upper bound on $\Lambda_*$, beyond which the theory requires further UV completion. In this case, a St\"uckelberg limit does not exist, as we elaborate on in section \ref{sec:nostuck}. %In \ref{sec:U13}, we discuss the implications of our results for theories featuring $U(1)^3$ anomalies at low energies.

%%%%%%%%%%%%%%%%%%%%%%%%%%%%%%%%%%%%%%%%%%%%%%%%%%
\subsection{Chiral lagrangian analysis}
\label{sec:EFT}

Although the irrelevant operators appearing in Eq.(\ref{Eq:fermionmass}) are the minimal set required to preserve gauge invariance, the symmetries allow (and one in general expects) a whole host of irrelevant operators to appear, any of which could point to the scale $\Lambda_*$ at which the anomaly-free EFT breaks down. While these could be enumerated by simply writing down the most general set of gauge-invariant irrelevant operators involving $A_\mu, \theta, \chi$, and $\psi$, this does not provide clear guidance as to the relative size of the various operators, and hence to the size of the corresponding cutoff $\Lambda_*$.

Indeed, not all irrelevant operators in the anomaly-free EFT are created equal. This can be seen most clearly by considering the simplest UV completion of the anomaly-free EFT: the Abelian Higgs model. The operators in Eq.(\ref{Eq:fermionmass}) can be obtained from a theory of a massless vector, massless fermions, and complex scalar $\Phi$ with Lagrangian
\begin{eqnarray}
	\mathcal{L} = |D_\mu \Phi|^2 - \frac{\lambda}{2} \left( |\Phi|^2 - \frac{f^2}{2} \right)^2 -  \left( y \Phi \bar \psi_L \psi_R + {\rm h.c.} \right) + \mathcal{L}(A_\mu, \psi, \chi) \ ,
\end{eqnarray}
where $D_\mu \Phi = (\partial_\mu - i (g_L - g_R) A_\mu) \Phi $. The potential for $\Phi$ leads to spontaneous symmetry breaking, which can be conveniently parameterized in terms of a radial mode $\rho$ and goldstone $\theta$ via
\begin{equation}
	\Phi = \frac{1}{\sqrt{2}} \left( f + \rho \right) e^{\frac{i \theta}{f}} \,.
\end{equation}
The radial mode $\rho$ acquires a mass $m_\rho = \sqrt{\lambda} f$ and may be integrated out to give an effective lagrangian of the form
%\begin{equation}
%	\mathcal{L}_{\rm eff} = \frac{1}{2} (\partial \theta)^2 + \frac{1}{2 f^2 (\lambda f^2)} (\partial \theta)^4  - M_f e^{i \theta /f} \bar \psi_L \psi_R + \frac{M_f}{f^2 (\lambda f^2)} (\partial \theta)^2 \bar \psi_L \psi_R + {\rm h.c.} + \dots
%\end{equation}
\begin{equation} \begin{aligned}
	\mathcal{L}_{\rm eff} =	& \frac{1}{2} (\partial \theta)^2 + \frac{1}{2 f^2 (\lambda f^2)} (\partial \theta)^4  \\
						& - \left( M_f e^{ \frac{i \theta}{f} } \bar \psi_L \psi_R - \frac{M_f}{f^2 (\lambda f^2)} (\partial \theta)^2 \bar \psi_L \psi_R \right) + {\rm h.c.} + \dots
\end{aligned} \end{equation}
This contains precisely the effective operator appearing in Eq.(\ref{Eq:fermionmass}), as well as higher-derivative terms for $\theta$, and derivative couplings between $\theta$ and the fermions. The former operator arose even before integrating out $\rho$ and is correspondingly independent of $m_\rho$, while the latter are generated by integrating out $\rho$ and proportional to negative powers of $m_\rho$. While this is suggestive, one would like to understand the natural size of various operator coefficients independent of the specific UV completion.

The relative size of these irrelevant operators can be understood more generally by approaching the anomaly-free EFT from the perspective of the chiral symmetries weakly gauged by the vector field and broken by the fermion mass, allowing us to bring NDA \cite{Manohar} power counting to bear on the problem. To do so, we begin by framing the non-anomalous EFT as one in which the $U(1)_L \times U(1)_R$ chiral symmetry of $\psi_L, \psi_R$ is weakly gauged and spontaneously broken. Under the $U(1)_L \times U(1)_R$ global symmetry the fermion fields transform as $\psi_L \rightarrow L \psi_L$ and $\psi_R \rightarrow R \psi_R$. Parameterizing $L,R$ as $L = e^{i (\alpha + \beta) / 2}$ and $R = e^{i ( - \alpha + \beta) / 2} $ for real $\alpha, \beta$, we have axial transformations $L R^\dag = e^{ i \alpha}$. The fermion mass arises due to unspecified (and potentially strong) interactions that break $U(1)_L \times U(1)_R \rightarrow U(1)_V$, giving one goldstone mode, which we can organize as
\begin{equation}
U = e^\frac{i \pi}{f} \ ,
\end{equation}
where $U$ transforms linearly under $U(1)_L \times U(1)_R$, $U \rightarrow U^\prime = L U R^\dag$. The goldstone $\pi$ correspondingly transforms under the shift $\pi \rightarrow \pi^\prime = \pi +  \alpha f$. (Although we will identify $\pi$ with $\theta$ momentarily, it is useful to differentiate the two for the time being.) From this perspective, the Abelian gauge symmetry can be thought of as gauging a particular subgroup of the vector and axial chiral symmetries, under which $\alpha = (Q_L - Q_R) \gamma$ and $\beta = (Q_L + Q_R) \gamma$. Clearly when $Q_L = Q_R$ we are gauging the vector symmetry preserved by the fermion mass, while for $Q_L \neq Q_R$ chiral symmetry breaking necessarily implies gauge symmetry breaking. 

We can then construct the non-anomalous EFT as the most general one invariant under the local $U(1)_L \times U(1)_R$ symmetries. The leading derivative interaction allowed by these symmetries is 
\begin{eqnarray}
\mathcal{L} \supset \frac{f^2}{2} \left| D_\mu U \right|^2  
= \frac{1}{2} (\partial \pi)^2 - f(g_L - g_R) A_\mu \partial^\mu \pi + \frac{1}{2} f^2 (g_L - g_R)^2  A_\mu A^\mu \ .
\end{eqnarray}
From this, it is clear that the chiral lagrangian can be matched to the anomaly-free EFT by making the identifications $\pi = \theta$ and $f | g_L - g_R | = m_\gamma$. Note the latter identification is a consequence of starting with the chiral symmetry breaking --- in reality, it is possible for the gauge symmetry to be broken more strongly than the chiral symmetry (by e.g.~a UV completion in which multiple scalars acquire vevs and contribute to the mass of the vector, while only one scalar couples to the fermions), but not visa versa. So one expects in full generality $m_\gamma \geq | g_L - g_R | f$ whenever $f$ denotes the scale of chiral symmetry breaking.

The chiral lagrangian formulation then allows us to use NDA power counting to enumerate irrelevant operators consistent with the symmetries and estimate the size of the corresponding Wilson coefficients in terms of their dependence on ${\bar g}, \Lambda_*,$ and $f$, where ${\bar g}$ is defined implicitly through the relation $\Lambda_* = {\bar g} f$. The most interesting operators for our purposes include (with Hermitian conjugates added where appropriate)
\begin{eqnarray}
\mathcal{O}_{y} &=& c_y M_f U^\dag \bar \psi_R \psi_L \ , \\
\mathcal{O}_{n} &=& \frac{c_{n}}{{\bar g}^2 \Lambda_*^{2n-4}} |D_\mu U|^{2n} \ , \\
\mathcal{O}_{f} &=& \frac{c_{f} M_f}{\Lambda_*^2}  (D_\mu D^\mu U^\dag) \bar \psi_R \psi_L \ , \ {\rm and}\\
\mathcal{O}_{L,R} &=& c_{L,R} U^\dag D_\mu U J_{L,R}^\mu  \ ,
\end{eqnarray}
where the coefficients $c_i$ are all $\mathcal{O}(1)$ numbers. In general, the precise values of the $c_i$ are not fixed and depend on details of the UV completion. However, there is one exception: we {\it must} have $|c_y| = 1$, since $\mathcal{O}_y$ defines the fermion mass $M_f$. Indeed, we recognize $\mathcal{O}_y$ (plus its Hermitian conjugate) as giving the operators in Eq.(\ref{Eq:fermionmass}), now reproduced via the chiral Lagrangian.

As for the remaining factors, NDA power counting cleanly distinguishes different operator classes. The operator $\mathcal{O}_y$ is independent of ${\bar g}$ and depends only on $M_f$ and $f$ (via $U$). In contrast, operators such as $\mathcal{O}_{n}$ and $\mathcal{O}_{f}$ depend on ${\bar g}$ via $\Lambda_*$. This agrees precisely with the Abelian Higgs UV completion considered earlier, which corresponds to $\lambda = {\bar g}^2$, $c_y = -1, c_{2} = \frac{1}{2},c_{f} = -1$, and $c_{L,R} = 0$.

In principle, any of these operators can lead to an upper bound on the scale $\Lambda_*$ at which the anomaly-free EFT breaks down. For example, the operators $\mathcal{O}_n$ give the leading unitarity violation at large Mandelstam $s$, independent of $M_f$, implying $\Lambda_* \sim 4 \pi f / \sqrt{c_n}$. Taking $M_f \rightarrow 0$ only decouples some of the irrelevant operators allowed by the symmetries of the non-anomalous EFT. This leads to a refinement of our criteria for acheiving the St\"uckelberg limit for arbitrary $m_\gamma$: (\emph{i}) when $M_f \rightarrow 0$ {\it and} either $f \rightarrow \infty$ or $c_i \rightarrow 0$ $\forall \ i$, regardless of the left- and right-handed charge assignments, and (\emph{ii}) when $Q_R \rightarrow Q_L$, regardless of $M_f, f,$ and the coefficients $c_i$. 

In practice, however, when setting a {\it quantitative} upper bound on the scale $\Lambda_*$ it bears emphasizing that only the coefficient of $\mathcal{O}_y$ is uniquely fixed in terms of the spectrum of the anomaly-free EFT, whereas the coefficients of all other irrelevant operators depend on the UV completion. This is in contrast to massive {\it non-Abelian} theories, where the same two-derivative operator that gives rise to the goldstone kinetic terms also induces derivative interactions with fixed relative coefficients that can be used to bound the cutoff. As such, only $\mathcal{O}_y$ leads to a model-independent bound on $\Lambda_*$ in the Abelian case. As for the other operators, one might imagine a UV completion in which all of the $c_i$ except $c_y$ are suppressed (or tuned) to approach a St\"uckelberg limit. This justifies proceeding with an analysis of the breakdown of perturbative unitarity using $\mathcal{O}_y$, or, equivalently the operators in Eq.(\ref{Eq:fermionmass}), to which we turn next.

%%%%%%%%%%%%%%%%%%%%%%%%%%%%%%%%%%%%%%%%%%%%%%%%%%
\subsection{Loss of perturbative unitarity}
\label{sec:tree}

To set a model-independent upper bound on $\Lambda_*$ in the anomaly-free EFT, we will study the high-energy behaviour of scattering amplitudes in a theory described by the following lagrangian:
\begin{equation}
\begin{split} \label{eq:fulltheory}
	\mathcal{L} = 	& - \frac{1}{4} F_{\mu \nu} F^{\mu \nu} + \frac{1}{2} (\partial_\mu \theta - m_\gamma A_\mu)^2 \\
				& + \bar \psi i \gamma^\mu \partial_\mu \psi + g_L A_\mu J^\mu_L + g_R A_\mu J^\mu_R - M_f \left[ \cos \left( \frac{\theta}{f} \right) \bar \psi \psi \pm i \sin \left( \frac{\theta}{f} \right) \bar \psi \gamma^5 \psi \right] \\
				& + \mathcal{L}_{\rm g.f.} + \mathcal{L}_{\chi} \ ,
\end{split}
\end{equation}
where $\mathcal{L}_{\rm g.f.} = - \frac{1}{2\xi} \left( \partial_\mu A^\mu + \xi m_\gamma \theta \right)^2$ is a standard gauge-fixing term, $J^\mu_{L,R}$ is as given in Eq.(\ref{eq:Jpsi}), and $\mathcal{L}_{\chi}$ refers to the terms in the lagrangian involving massless fermions (kinetic terms plus couplings to $A_\mu$), which will not be relevant for the subsequent discussion, other than noting that their charge assignments are such that the full theory is anomaly-free. (As in Eq.(\ref{Eq:fermionmass}), the $+$ ($-$) sign corresponds to the case $Q_L - Q_R > 0$ ($Q_L - Q_R < 0$).)

\begin{figure}
\centering
  \includegraphics[scale=1.2]{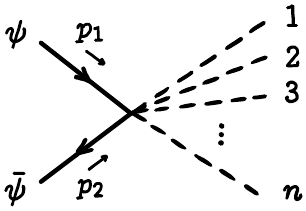}
  \captionof{figure}{Eq.(\ref{Eq:fermionmass}) contains operators coupling two fermions to any number of $\theta$'s. At high center of mass energies, this contact interaction provides the leading contribution to the scattering process $\psi \bar \psi \rightarrow n \theta$.}
  \label{fig:tree}
\end{figure}

A process that reflects the breakdown of perturbation theory in the context of the EFT described by Eq.(\ref{eq:fulltheory}) concerns scattering of a fermion--anti-fermion pair into a number of longitudinal vectors at high center of mass energy. Making use of the Goldstone equivalence theorem, we can obtain the leading high-energy contribution to the corresponding scattering amplitude by considering the process $\psi \bar \psi \rightarrow n \theta$ in a general gauge. Expanding Eq.(\ref{Eq:fermionmass}) as a power series in $\theta / f$, we see that it contains operators coupling two fermions to any number of $\theta$'s:
\begin{equation}
	\mathcal{L} \supset - M_f \left[ \bar \psi \psi \sum_{ \{n \ {\rm even} \} } \frac{(-1)^{\frac{n}{2}}}{n!} \left( \frac{\theta}{f}\right)^n \pm i \bar \psi \gamma^5 \psi \sum_{ \{ n \ {\rm odd} \} } \frac{(-1)^{\frac{n-1}{2}}}{n!} \left( \frac{\theta}{f}\right)^n \right] \ .
\end{equation}
At high energies, $\sqrt{s} \gtrsim M_f$, the leading contribution is due to the contact operator depicted in Figure~\ref{fig:tree} --- diagrams involving more vertices will contain further powers of $M_f$, and therefore feature a milder high-energy behaviour.
%%%%%%%%%%%%%%%%%%%%
%%%%%%%%%%%%%%%%%%%%

The observation that the presence of massive fermions with chiral charge assignments leads to the breakdown of perturbation theory at sufficiently high scales was first made in \cite{Appelquist:1987cf} in the context of the Standard Model without a dynamical mechanism responsible for generating fermion masses, and later refined in \cite{Maltoni:2001dc,Dicus:2004rg}. Of course, in the `Higgsless' Standard Model, a bound $\Lambda \lesssim 4 \pi v$ already follows from the loss of perturbative unitarity in longitudinal gauge boson scattering. Thus, for a non-zero fermion mass to reliably indicate a cutoff scale parametrically above the weak scale, the mechanism responsible for restoring perturbation theory at scales above $\sim 4 \pi v$  needs to be introduced into the analysis. In contrast, for a massive $U(1)$, the two-derivative terms in the action involving only longitudinal gauge bosons do not imply the loss of perturbative unitarity. Although higher-derivative operators such as the $\mathcal{O}_n$ encountered in section \ref{sec:EFT} {\it can} lead to the breakdown of perturbation theory in longitudinal gauge boson scattering, the precise bound in this case depends on the UV-completion via the unknown operator coefficients $c_n$. We can, however, derive a model-independent upper bound on the cutoff scale that stems purely from the presence of massive chiral fermions. 

As discussed in \cite{Falkowski:2019tft}, a reasonable estimate of the range of validity of perturbation theory within a unitary theory can be obtained by demanding that
\begin{equation} \label{eq:generalubound}
	%\langle \Psi | T^\dagger T | \Psi \rangle = \sum_X | \langle X | T | \Psi \rangle |^2 \lesssim \pi^2 \ ,
	\langle \Psi | T^\dagger T | \Psi \rangle  = \sum_X | \mathcal{M} ( \Psi \rightarrow X ) |^2 \lesssim \pi^2 \ ,
\end{equation}
for any unit-normalized state $| \Psi \rangle$, and where $i T = S - \mathbb{1}$ as usual. $\sum_X$ refers to a sum over all possible final states, with an integral over the corresponding phase space being implicit. For the case at hand, we will estimate the scale of perturbation theory breakdown when the following inequality is saturated:
\begin{equation} \label{eq:ubound}
	\sum_{n=2}^\infty \frac{1}{n!} \left. \int_{\Pi_n} | \mathcal{M} ( \psi {\bar \psi} \rightarrow n \theta ) |^2 \right|_{\sqrt{s} = \Lambda_*} \lesssim \pi^2 \ ,
\end{equation}
where $\int_{\Pi_n}$ refers to the integral over the final $n$-body phase space, and the factor of $1/n!$ takes care of the fact that all $n$ particles in the final state are identical. The scale $\Lambda_*$ provides an upper bound on the cutoff scale at which the theory requires UV-completion.

Allowing for the possibility of several fermion species with the same value of $|Q_L - Q_R|$, we choose our initial state to be the spin-singlet:
\begin{equation} \label{eq:initialstate}
	| \psi \bar \psi \rangle \equiv \frac{1}{\sqrt{2 N}} \sum_{a=1}^N | \psi^a_+ {\bar \psi}^a_+ - \psi^a_- {\bar \psi}^a_- \rangle \ .
\end{equation}
This specific choice is of course not necessary to derive a unitarity bound. However, choosing the initial state to be a spin-eigenstate is convenient since in that case only operators with $n$ being either even or odd lead to a non-vanishing contribution, simplifying our analysis. Including a (conveniently normalized) sum over fermion flavors allows us to keep track of factors of $N$, which may be relevant when $N \gg 1$.

Further specifying our initial state to be the $s$-wave component of Eq.(\ref{eq:initialstate}), we have, for even $n$: 
\footnote{As advertised, to use a bound of the form of Eq.(\ref{eq:ubound}), the initial state must be unit-normalized. This can be achieved by choosing the initial state to be a spherical wave, which for an $s$-wave leads to the corresponding scattering amplitude being smaller by a factor of $\sqrt{16 \pi}$ compared to a plane-wave calculation.}
\begin{equation} \begin{split}
	i \mathcal{M} ( \psi {\bar \psi} \rightarrow n \theta )
				& = - \frac{i}{\sqrt{16 \pi}} \frac{(-1)^\frac{n}{2}}{\sqrt{2N}} \frac{M_f}{f^n} \sum_{a=1}^N \left[ {\bar v^a}_+ (p_2) u^a_+ (p_1) - {\bar v^a}_- (p_2) u^a_- (p_1) \right] \\
				&  \simeq - \frac{i}{\sqrt{16 \pi}} (-1)^\frac{n}{2} \sqrt{2N} \frac{M_f}{f^n} \sqrt{s} \ ,
\end{split} \end{equation}
where we have used ${\bar v^a}_\pm (p_2) u^a_\pm (p_1) \simeq \pm \sqrt{s}$ at large momentum. The asymptotic expression for the volume of the $n$-body phase space factor when all particles in the final state are massless is given by \cite{Kleiss:1985gy,Dicus:2004rg}
\begin{equation}
	\int_{\Pi_n} \simeq \frac{2 \pi}{(4 \pi)^{2(n-1)}} \frac{s^{n-2}}{(n-1)! (n-2)!} \ .
\end{equation}
Up to overall $\mathcal{O}(1)$ corrections, Eq.(\ref{eq:ubound}) can then be written as
\begin{equation} \label{eq:uboundv2}
	\sum_{n=2}^\infty \frac{1}{n! (n-1)! (n-2)!} \left( \frac{\Lambda_*}{4 \pi f} \right)^{2(n-1)} \equiv \mathcal{F} \left( \frac{\Lambda_*}{4 \pi f} \right) \lesssim \left( \frac{ 4 \pi f}{\sqrt{N} M_f} \right)^2 \ ,
\end{equation}
where $\mathcal{F} (x) \equiv \frac{x^2}{2} {}_0F_5 (\frac{1}{2}, 1, \frac{3}{2}, \frac{3}{2}, 2; \frac{x^4}{64})$, and ${}_0F_q ({b_1, \cdots, b_q} ; z)$ is a generalized hypergeometric function. We do not need the specific form of $\mathcal{F} (x)$, except for noting that it is a monotonically increasing function, with asymptotic expansions at small and large $x$ (up to irrelevant overall $\mathcal{O} (1)$ factors):
\begin{align}
\mathcal{F} (x) \sim
		\begin{dcases}
		x^2 \quad & {\rm for} \ x \lesssim 1 \\
		\frac{e^{3 x^{2/3}}}{x^{2/3}} \quad & {\rm for} \ x \gtrsim 1 \\
                	%x^2 \quad \col{ \frac{x^2}{2} } & {\rm for} \ x \lesssim 1 \\
                	% \frac{e^{3 x^{2/3}}}{x^{2/3}} \quad \col{ \frac{e^{3 x^{2/3}}}{4 \sqrt{3} \pi x^{2/3}} } & {\rm for} \ x \gtrsim 1 \\
                	\end{dcases}
\end{align}

In the regime where $M_f \lesssim 4 \pi f / \sqrt{N}$, the right-hand-side of Eq.(\ref{eq:uboundv2}) is always $\gtrsim 1$, and so it is appropriate to expand $\mathcal{F}(x)$ for $x\gtrsim 1$. Eq.(\ref{eq:uboundv2}) then reads
\begin{equation}
	\left. \frac{e^{3 x^{2/3}}}{x^{2/3}} \right|_{x= \frac{\Lambda_*}{4 \pi f}} \lesssim \left( \frac{4 \pi f}{\sqrt{N} M_f} \right)^2 \ .
\end{equation}
Thus, up to $\mathcal{O}(1)$ corrections, we have
\begin{equation} \label{eq:Lambda*}
	\boxed{	\Lambda_* \lesssim 4 \pi f \left( \log \frac{4 \pi f}{\sqrt{N} M_f} \right)^{3/2} = \frac{4 \pi m_\gamma}{|g_L - g_R| } \left( \log \frac{4 \pi m_\gamma}{\sqrt{N} M_f |g_L - g_R|} \right)^{3/2} } \ ,
\end{equation}
where in the last step we have written $f$ in terms of the vector mass, and the left- and right-handed gauge couplings (remember Eq.(\ref{eq:f})). Eq.(\ref{eq:Lambda*}) provides a model-independent upper bound on the cutoff scale of a massive Abelian gauge theory coupled to ($N$ copies of) a massive fermion featuring chiral charge assignments, and it is the main result of this section. 

An a priori weaker bound could be obtained by applying the perturbativity bound to each term in Eq.(\ref{eq:ubound}), instead of performing the full sum over $n$, and this highlights how different values of $n$ dominate the overall bound in the different regimes. For instance, the case $n=2$ already implies a non-trivial bound:
\begin{equation}
	\Lambda_*^{(n=2)} \lesssim 4 \pi f \frac{4 \pi f}{\sqrt{N} M_f} \ ,
\end{equation}
which agrees with Eq.(\ref{eq:Lambda*}) when $M_f \sim 4 \pi f / \sqrt{N}$, but leads to a much weaker bound in the limit of small fermion mass $M_f \ll 4 \pi f / \sqrt{N}$. In general, we may impose the perturbativity bound on the $n$-th term in Eq.(\ref{eq:ubound}). After expanding for large values of $n$, using Stirling's approximation $n! \simeq \left( \frac{n}{e} \right)^n \sqrt{2 \pi n}$, we find
\begin{equation} \label{eq:Lambda*n}
	\Lambda_*^{(n)} \lesssim 4 \pi f \left( \frac{n-1}{e}\right)^\frac{3}{2} \left\{ \left( 2 \pi (n-1) \right)^{3/2} \left( \frac{4 \pi f}{\sqrt{N} M_f} \right)^2 \right\}^\frac{1}{2(n-1)} \ .
\end{equation}
The value of $n$ for which the bound is the strongest is given by
\begin{equation}
	n_* - 1\simeq \frac{2}{3} \log \frac{4 \pi f}{\sqrt{N} M_f} \sim  \log \frac{4 \pi f}{\sqrt{N} M_f} \ ,
\end{equation}
which is large for $M_f \ll 4 \pi f / \sqrt{N}$, justifying our earlier approximation. Evaluating Eq.(\ref{eq:Lambda*n}) for $n=n_*$, we find
\begin{equation}
	\Lambda_*^{(n=n_*)} \lesssim 4 \pi f \left( \log \frac{4 \pi f}{\sqrt{N} M_f} \right)^{3/2} \ ,
\end{equation}
which reproduces Eq.(\ref{eq:Lambda*}).

Our bound Eq.(\ref{eq:Lambda*}) applies so long as all massive fermion species appear roughly at the same scale $M_f$, and the value of $q_i \equiv | Q_L^{(i)} - Q_R^{(i)} |$ is parametrically of the same size for all $i = 1, \cdots, N$. If the $q_i$ are parametrically different, then the bound will be dominated by the species with the largest $q_i = q_{\rm max}$. In this case, the strongest bound can be obtained by choosing an initial state that involves only the fermion with the largest $q_i$ (instead of summing over flavors, as in Eq.(\ref{eq:initialstate})), and the resulting bound is just Eq.(\ref{eq:Lambda*}) with $N \rightarrow 1$, and $f$ as given in Eq.(\ref{eq:f}) with $| Q_L - Q_R | \rightarrow q_{\rm max}$.

%%%%%%%%%%%%%%%%%%%%%%%%%%%%%%%%%%%%%%%%%%%%%%%%%%
\subsection{No St\"uckelberg limit}
\label{sec:nostuck}

\begin{figure}
\centering
  \includegraphics[scale=1.2]{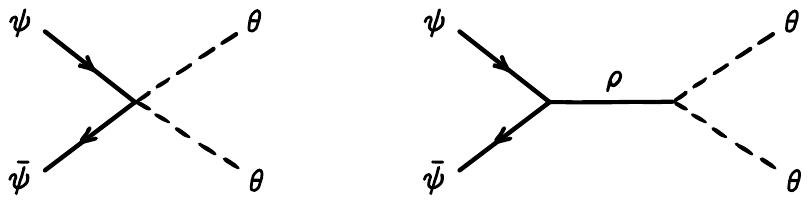}
  \captionof{figure}{The presence of a radial mode featuring derivative couplings to $\theta$ is necessary in order to cancel the pathological high energy behaviour in $\psi \bar \psi \rightarrow n \theta$. When $n=2$, the diagram on the right precisely cancels the leading high energy piece of the left graph.}
  \label{fig:h}
\end{figure}

Eq.(\ref{eq:Lambda*}) provides a finite, model-independent upper bound on the cutoff scale of an Abelian gauge theory that contains massive chiral fermions. As advertised in the Introduction, no St\"uckelberg limit exists in this class of models, and the theory requires further UV-completion at or below the scale $\Lambda_*$ in order to recover a perturbative expansion. If this UV-completion is in the form of a weakly coupled Abelian Higgs model, $\Lambda_*$ is an upper bound on the mass of the radial mode, $\rho$, and it cannot be decoupled. Due to the derivative couplings between $\rho$ and the longitudinal mode of the massive vector, of the form $\mathcal{L} \supset \frac{\rho}{f} (\partial_\mu \theta)^2$, diagrams involving the radial mode now cancel the pathological high energy growth of scattering amplitudes describing the process $\psi \bar \psi \rightarrow n \theta$. This is illustrated in Figure~\ref{fig:h} for $n=2$.
%%%%%%%%%%%%%%%%%%%%
%%%%%%%%%%%%%%%%%%%%

The behavior of Eq.(\ref{eq:Lambda*}) is also consistent with our expectations: the upper bound on the cutoff decouples in the limit where the theory contains no massive chiral fermions. This can happen in two ways: in the limit $| Q_L - Q_R | \rightarrow 0$, regardless of the fermion mass, and in the limit $M_f \rightarrow 0$, regardless of the left- and right-handed charges. In the former case, all massive fermions couple to the gauge field through a vector current, the theory is anomaly-free at all scales, and, in keeping with the discussion of section~\ref{sec:intro}, no model-independent upper bound exists on the scale of UV-completion. Similarly, the limit $M_f \rightarrow 0$ also allows for a St\"uckelberg limit, but the cutoff scale decouples much slower than before. Rewriting Eq.(\ref{eq:Lambda*}) as:
\begin{equation}
	M_f \sim \frac{4 \pi f}{\sqrt{N}} e^{- \left( \frac{\Lambda_*}{4 \pi f} \right)^{2/3}} \ ,
\end{equation}
it becomes apparent that a cutoff scale that is parametrically above $4 \pi f$ requires the massive chiral fermions to appear exponentially below this scale. As noted in section~\ref{sec:EFT}, in this case the existence of other UV-dependent irrelevant operators whose coefficients are independent of $M_f$ additionally requires $c_i \rightarrow 0$ $\forall \ i$ to prevent them from independently obstructing the St\"uckelberg limit.

Barring the possibility of massive chiral fermions that are exponentially light, we may ignore the $\log$ factor in Eq.(\ref{eq:Lambda*}). In this case, the upper bound on the scale of UV-completion is parametrically given by
\begin{equation}
\label{eq:Lambdaup}
	\Lambda_* \lesssim \frac{4 \pi m_\gamma}{|g_L - g_R|} = \frac{4 \pi m_\gamma}{g_0 |Q_L - Q_R|} \ ,
\end{equation}
which can be rewritten as a lower bound on the photon mass:
\begin{equation}
\label{eq:mgammalow}
	m_\gamma \gtrsim \frac{|g_L - g_R| \Lambda_*}{4 \pi} = \frac{g_0 |Q_L - Q_R| \Lambda_*}{4 \pi} \ .
\end{equation}

Thus, although an arbitrarily light vector is not possible in this class of models, in keeping with Eq.(\ref{eq:mgammalow}), a parametric separation of scales $m_\gamma \ll g \Lambda_* / 4 \pi$ remains a possibility so long as an equally large ratio of electric charges is introduced in the UV-completion:
\begin{equation}
	\frac{g}{ |g_L - g_R| } = \frac{Q}{ |Q_L - Q_R| } \gg 1 \ .
\end{equation}
In other words, the typical value of the gauge coupling present in the low energy theory must be much larger than the quantum of electric charge. Such a large ratio of charges allows us to probe the regime of a parametrically small vector mass that is only available to Abelian gauge theories (above the dotted line in Figure~\ref{fig:cutoff}). In the next section, we discuss the implications of introducing a large hierarchy of charges in theories where the low energy spectrum contains uncanceled gauge anomalies. If the anomalous EFT features a mixed gravitational anomaly, such a large ratio of charges will come at a hefty price.

%%%%%%%%%%%%%%%%%%%%%%%%%%%%%%%%%%%%%%%%%%%%%%%%%%
%%%%%%%%%%%%%%%%%%%%%%%%%%%%%%%%%%%%%%%%%%%%%%%%%%
\section{Abelian anomalies and very light vectors}
\label{sec:grav}

Consistent, four-dimensional gauge theories with anomalous fermion content in the infrared necessarily feature massive chiral fermions responsible for canceling the anomalies of the low energy spectrum. They are correspondingly subject to the general constraints obtained in section \ref{sec:unitarity}. In this section, we discuss the implications of these constraints for Abelian gauge theories that appear anomalous below a certain scale. In particular, we focus on the consequences of probing the regime of a parametrically light vector, which in turn requires the presence of a parametrically large ratio of electric charges, as discussed in section~\ref{sec:nostuck}.

We begin with purely non-gravitational phenomena, focusing on the effect of the $U(1)^3$ anomaly in section \ref{sec:U13}. We turn on gravity in section \ref{sec:gravonly}, and discuss the implications of a low energy fermion spectrum featuring a mixed $U(1)$-gravitational anomaly. Finally, in \ref{sec:both}, we consider the more generic case where both anomalies are present, and elaborate on their surprising interplay.

%%%%%%%%%%%%%%%%%%%%%%%%%%%%%%%%%%%%%%%%%%%%%%%%%%
\subsection{$U(1)^3$ anomaly}
\label{sec:U13}

Picking up where we left off in section \ref{sec:review}, we focus first on an anomalous EFT that contains a single massless fermion coupling to $A_\mu$ through a left-handed current. As advertised, we will first neglect gravitational interactions in our discussion, i.e.~we ignore the presence of a mixed gravitational anomaly, as well as any potential extra requirements stemming from quantum gravity consistency such as compactness of the gauge group.

In this case, the $U(1)^3$ anomaly of the anomalous EFT can be cancelled by introducing a single massive fermion with chiral charge assignments $Q_L$ and $Q_R$ such that
\begin{equation} \label{eq:U13_LH}
	{\rm tr}_{\rm all} ( Q_i^3 ) = Q^3 + Q_L^3 - Q_R^3 = 0 \ .
\end{equation}
As per Fermat's Last Theorem, the only integer solutions to this equation are either $\{ Q_R = Q, Q_L=0\}$ or $\{ Q_R = 0, Q_L=-Q \}$. However, since we are ignoring gravitational effects for the time being, we will allow ourselves to entertain non-integer solutions to Eq.(\ref{eq:U13_LH}), and WLOG we set $q \equiv Q_R - Q_L = 1$ in what follows. \footnote{Factors of $q$ may be restored by performing the simultaneous rescaling $Q_i \rightarrow Q_i / q$ and $g_0 \rightarrow g_0 q$.}

So long as the heavy fermion is not exponentially light, an upper bound on the cutoff scale of this anomaly-free extension is given by Eq.(\ref{eq:Lambdaup}):
\begin{equation} \label{eq:cutoff_LH}
	\Lambda_* \lesssim \frac{4 \pi m_\gamma}{g_0} = \frac{4 \pi m_\gamma}{g} Q \ .
\end{equation}
Probing the regime $\Lambda_* \gg 4 \pi m_\gamma / g$ therefore requires $Q \gg 1$, and a solution to Eq.(\ref{eq:U13_LH}) in this limit implies $Q_L \simeq Q^{3/2} / \sqrt{3}$. Taking $Q$ large, we can then push $\Lambda_*$, as well as the mass of the heavy fermion, parametrically above $\sim 4 \pi m_\gamma / g$. An upper bound on $Q$, however, stems from the requirement that the heavy fermion remains weakly coupled, parametrically:
\begin{equation} \label{eq:U13_weak}
	\frac{g_0^2 \left( Q_L^2 + Q_R^2 \right)}{16 \pi^2} \sim \frac{g_0^2 Q^3}{ 16 \pi^2 } \lesssim 1 \qquad \Rightarrow \qquad Q \lesssim \frac{16 \pi^2}{g^2} \ .
\end{equation}
With this requirement, we obtain an upper bound on the scale of UV-completion, of the form
\begin{equation} 
	\Lambda_* \lesssim \frac{64 \pi^3 m_\gamma}{g^3} \ ,
\end{equation}
which coincides with the scale $\Lambda_{U(1)^3}$ in Eq.(\ref{eq:Lambda3}). Since $M_f \lesssim \Lambda_*$, this allows us to saturate the upper bound on the cutoff of the anomalous EFT obtained in \cite{Preskill:1990fr}, at the cost of a further UV-completion incorporating a radial mode appearing roughly at the same scale. 

Although we have focused the discussion on an anomalous EFT with a single left-handed fermion, our conclusion applies more generally in the context of anomalous EFTs with $U(1)^3$ anomalies so long as the  infrared fermion spectrum does not feature wild hierarchies of charges. Even with gravity turned on, the above result applies, parametrically, so long as ${\rm tr}_{\rm IR} (Q_i) = 0$, and ${\rm tr}_{\rm IR} (Q_i^3) \sim Q^3$, with $Q$ the typical charge present in the low energy spectrum. A concrete example is a theory with massless left-handed fermions carrying charges $Q$, $Q$, and $-2 Q$. This theory has a vanishing $U(1)$-gravitational anomaly, whereas ${\rm tr}_{\rm IR} (Q_i^3) = - 6 Q^3 \sim Q^3 $. We can extend this theory into one free of anomalies by introducing two pairs of massive chiral fermions, with charges:
\begin{equation}
	Q_R^{(1)} - Q_L^{(1)} = - \left(Q_R^{(2)} - Q_L^{(2)} \right) \equiv 1 \ .
\end{equation}
(As before, we have set the right-hand-side above to unity WLOG.) With this charge assignments, the heavy fermions do not introduce a mixed gravitational anomaly. The upper bound in Eq.(\ref{eq:cutoff_LH}) similarly applies in this case, and $Q \gg 1$ is required to achieve a parametrically high cutoff. As in our previous example, an upper bound on $Q$ stems from the requirement that the massive fermions remain weakly coupled, while at the same time having appropriate charge assignments so as to cancel the $U(1)^3$ anomaly. The optimal choice of charges, i.e.~allowing for the largest $Q$ while maintaining perturbativity, is such that $Q_L^{(2)} \gg Q_L^{(1)}$, in which case anomaly-cancelation in turn requires $Q_L^{(2)} \simeq \sqrt{2} Q^{3/2} \gg 1$. Moreover, unlike in our previous example, integer solutions to the anomaly equations now exist for values $Q \gg 1$. The resulting upper bound on $Q$ from the requirement of weak coupling is, again, given by Eq.(\ref{eq:U13_weak}), and therefore $\Lambda_* \lesssim 64 \pi^3 m_\gamma / g^3$ follows.

Thus, in general, Abelian gauge theories that only feature a $U(1)^3$ anomaly at low energies may be UV-completed into anomaly-free extensions by introducing an $\mathcal{O}(1)$ number of massive fermion species, while at the same time allowing for a parametrically large ratio of electric charges. In turn, this allows us to probe the regime $m_\gamma \ll g \Lambda_* / 4 \pi$, in keeping with the results of \cite{Preskill:1990fr}, while maintaining a quantum gravity cutoff at scales of order $M_{Pl}$.

%%%%%%%%%%%%%%%%%%%%%%%%%%%%%%%%%%%%%%%%%%%%%%%%%%
\subsection{Mixed $U(1)$-gravitational anomaly}
\label{sec:gravonly}

We now focus on the implications of a mixed $U(1)$-gravitational anomaly by considering theories such that ${\rm tr}_{\rm IR} (Q_i^3) = 0$ but ${\rm tr}_{\rm IR} (Q_i) \sim Q$, with $Q$ the typical charge of the low energy spectrum. Although such charge assignments may seem non-generic, the purpose of this section is to illustrate the effect of the mixed gravitational anomaly, without distractions stemming from additional requirements imposed by the presence of a $U(1)^3$ anomaly. A specific example of this kind is the `taxicab number' theory with massless left-handed fermions carrying charges $- Q, -12 Q, 9 Q$, and $10 Q$.

Cancelling the gravitational anomaly without introducing a $U(1)^3$ anomaly requires more than a single massive fermion. In general, we may introduce a number $N$ of chiral fermion species, with charge assignments
\begin{equation}
	Q_R^{(i)} - Q_L^{(i)} \equiv q_i \ .
\end{equation}
The requirement that the massive fermions cancel the mixed gravitational anomaly of the low energy spectrum can then be written as \footnote{For our taxicab number theory, Eq.(\ref{eq:gravanomaly}) reads $q_1 + \cdots + q_N = 6 Q \sim Q$.}
\begin{equation} \label{eq:gravanomaly}
	q_1 + \cdots + q_N = {\rm tr}_{\rm IR} (Q_i) \sim Q \ .
\end{equation}
If one, or an $\mathcal{O}(1)$ number, of the $q_i$ is much larger than the rest, then satisfying Eq.(\ref{eq:gravanomaly}) requires $q_{\rm max} \sim Q$.
As discussed in section~\ref{sec:nostuck}, the upper bound on the scale of UV-completion is then given by Eq.(\ref{eq:Lambda*}) with $N \sim 1$ and $|Q_L - Q_R| \sim q_{\rm max} \sim Q$. Ignoring the log factor, this leads to $\Lambda_* \lesssim 4 \pi m_\gamma / g$ --- i.e.~a parametric separation of scales is not possible if an $\mathcal{O}(1)$ number of heavy fermions is responsible for cancelling the anomaly of the low energy EFT. \emph{Instead, probing the regime of a parametrically light vector in a theory with a mixed gravitational anomaly in the infrared requires the presence of a parametrically large number of massive fermion species, all of which contribute significantly to the anomaly.} In turn, this implies a quantum gravity cutoff that lies parametrically below $M_{Pl}$.

The optimal charge assignment (that is, requiring the smallest number of species for a given $Q$) is such that the largest possible number of $q_i$ have equal sign, adding coherently in Eq.(\ref{eq:gravanomaly}) to cancel the anomaly of the low energy EFT. The case with $q_i = q$ $\forall$ $i=1, \cdots , N$ does not allow for the massive fermions to be themselves free of $U(1)^3$ anomalies. However, it will generally be enough to have  $q_i = q$ for an $\mathcal{O}(1)$ fraction of all the fermion species, with a small number of the $q_i$ having opposite sign. Setting $q=1$ WLOG, this implies \footnote{Again, for the taxicab number example, it is enough to have $q_i = 1$ for $i=1, \cdots, N-1$, and $q_N = - 1$. In this case, integer solutions to the anomaly equations exist, and Eq.(\ref{eq:gravanomaly}) reads $N - 2 = 6 Q \Rightarrow Q = (N - 2) / 6 \sim N$, in agreement with Eq.(\ref{eq:NsimQ}).}
\begin{equation}
\label{eq:NsimQ}
	N \sim Q \ .
\end{equation}
Thus, through the requirement that the $U(1)$-gravitational anomaly of the low energy spectrum is cancelled by the heavy fermions, the anomaly links a large ratio of charges, required to achieve a large separation of scales, to a large number of species.
Parametrically, the cutoff of our anomaly-free extension can now be written as
\begin{equation} \label{eq:Lambda*N}
	\Lambda_* \lesssim \frac{4 \pi m_\gamma}{g} Q \sim \frac{4 \pi m_\gamma}{g} N \ .
\end{equation}

An upper bound on $N$ immediately follows from the requirement that the scale of UV-completion falls bellow the quantum gravity scale, $\Lambda_{\rm QG} \sim 4 \pi {M}_{Pl} / \sqrt{N}$ \cite{Dvali:2007hz,Dvali:2007wp}. Demanding that $\Lambda_* \lesssim \Lambda_{\rm QG}$, we find
\begin{equation} \label{eq:Nmaxgrav}
	N \lesssim \left( \frac{g {M}_{Pl}}{m_\gamma} \right)^{2/3} \ .
\end{equation}
Plugging this back in Eq.(\ref{eq:Lambda*N}), we obtain an upper bound on the cutoff of the anomaly-free EFT, of the form
\begin{equation}
	\Lambda_* \lesssim 4 \pi \left( \frac{{M}_{Pl}^2 m_\gamma}{g} \right)^{1/3} \ ,
\end{equation}
which is precisely the upper bound on the cutoff scale of the anomalous EFT obtained in \cite{AlvarezGaume:1983ig,Preskill:1990fr}, given by $\Lambda_{\rm grav}$ in Eq.(\ref{eq:Lambdagrav}).

\begin{figure}
\centering
	\includegraphics[scale=1.2]{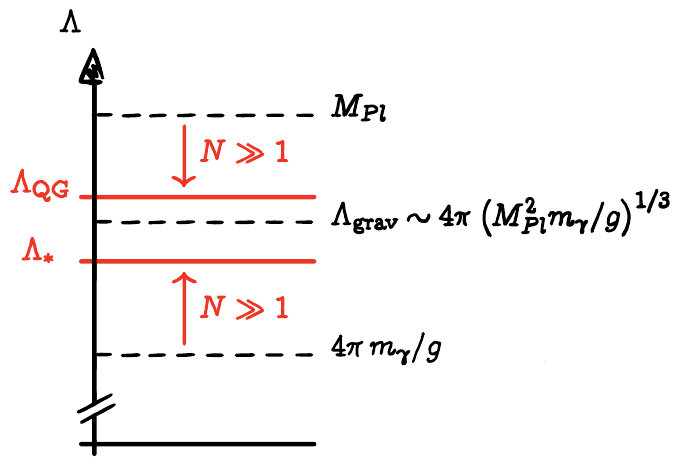}
 	\caption{The cutoff $\Lambda_*$ of an anomaly-free theory that UV-completes an anomalous EFT with a mixed $U(1)$-gravitational anomaly can be taken large at the cost of increasing the number $N$ of massive fermion species responsible for rendering the theory anomaly-free. In turn, this lowers the quantum gravity scale parametrically below $M_{Pl}$. The cutoff of the anomaly-free theory may be taken to saturate the apparent cutoff of the anomalous EFT, $\Lambda_{\rm grav}$, and all three scales parametrically coincide in that limit: $\Lambda_* \sim \Lambda_{\rm grav} \sim \Lambda_{\rm QG}$.}
\label{fig:grav}
\end{figure}

More generally, it is illuminating to obtain upper bounds on $\Lambda_*$ and $\Lambda_{\rm QG}$, which can be written in the following form
\begin{equation} \label{eq:upperbounds}
	\Lambda_* \lesssim 4 \pi \left( \frac{{M}_{Pl}^2 m_\gamma}{g} \right)^{1/3} \left( \frac{\Lambda_*}{\Lambda_{\rm QG}} \right)^{2/3} \qquad {\rm and} \qquad 
	\Lambda_{\rm QG} \lesssim 4 \pi \left( \frac{{M}_{Pl}^2 m_\gamma}{g} \right)^{1/3} \left( \frac{\Lambda_{\rm QG}}{\Lambda_*} \right)^{1/3} \ .
\end{equation}
Eq.(\ref{eq:upperbounds}) highlights how pushing the cutoff scale up comes at the cost of lowering the quantum gravity scale, as we further illustrate in Figure \ref{fig:grav}.
%%%%%%%%%%%%%%%%%%%%
%%%%%%%%%%%%%%%%%%%%
Our analysis further shows how the consistency requirement that an Abelian gauge theory that contains a $U(1)$-gravitational anomaly be UV-completed below the scale of Eq.(\ref{eq:Lambdagrav}) is self-imposed by the theory itself. As we push the scale of UV-completion up, the quantum gravity cutoff will correspondingly come down. In the limit where these two scales meet, they further coincide with the orange line on Figure~\ref{fig:cutoff}. In other words: it is not possible to access the Swampland of EFTs living in the orange region of Figure~\ref{fig:cutoff} from the subset of consistent theories lying below. More generally, our result provides evidence, purely in the context of field theory, of the existence of a Swampland of EFTs that are not smoothly connected to the subset of consistent theories that are compatible with a gravitational UV-completion. \footnote{See \cite{Draper:2019utz} for another example in a similar spirit.}

%%%%%%%%%%%%%%%%%%%%%%%%%%%%%%%%%%%%%%%%%%%%%%%%%%
\subsection{Both Abelian anomalies}
\label{sec:both}

We now turn to the more generic case where the low energy spectrum features both a $U(1)^3$ anomaly and a mixed gravitational anomaly. We will focus on the example of a theory with a massless left-handed fermion, but as before our conclusions will apply more generally so long as ${\rm tr}_{\rm IR} (Q_i^3) \sim Q^3$ and ${\rm tr}_{\rm IR} (Q_i) \sim Q$. Introducing $N$ massive fermion species such that
\begin{equation}
	Q_R^{(i)} - Q_L^{(i)} \equiv 1 \quad \forall \quad i = 1, \cdots , N
\end{equation}
is now enough for the heavy fermions to cancel both anomalies. Solving both the gravitational and $U(1)^3$ anomalies requires, respectively:
\begin{equation}
	Q = N \ , 
\end{equation}
\begin{equation}
	{\rm and} \qquad Q^3 = N \left( 3 Q_L^2 + 3 Q_L + 1 \right) \ ,
\end{equation}
where for simplicity we have assumed that $Q_L^{(i)} = Q_L$ $\forall \ i$. In the regime where $Q = N \gg 1$, as required to realize a large separation of scales, $Q_L \simeq N / \sqrt{3}$, and integer solutions exists for certain values of $N \gg 1$.

As in section \ref{sec:gravonly}, an upper bound on $N$ as given in Eq.(\ref{eq:Nmaxgrav}) follows from the requirement that the theory be UV-completed below the quantum gravity scale. However, the requirement that the heavy fermions be weakly coupled now sets an additional upper bound on $N$, of the form
\begin{equation} \label{eq:Nbound}
	\frac{g_0^2 N \left( Q_L^2 + Q_R^2 \right) }{16 \pi^2} \sim \frac{g_0^2 N^3}{16 \pi^2} \lesssim 1 \qquad \Rightarrow \qquad N \lesssim \left( \frac{16 \pi^2}{g_0^2} \right)^{1/3} \lesssim \frac{16 \pi^2}{g^2} \ ,
\end{equation}
which is just Eq.(\ref{eq:U13_weak}) after setting $Q = N$.

Comparing Eq.(\ref{eq:Nmaxgrav}) and Eq.(\ref{eq:Nbound}), we identify a critical value of the gauge coupling:
\begin{equation} \label{eq:Nmax3}
	g_* \sim \left( \frac{64 \pi^3 m_\gamma}{{M}_{Pl}} \right)^{1/4} \ .
\end{equation}
When $g \lesssim g_*$, the upper bound on Eq.(\ref{eq:Nmaxgrav}) is the most stringent, and the conclusions of section \ref{sec:gravonly} apply in this regime: $\Lambda_*$ can saturate the cutoff scale of the anomalous EFT, which in this regime corresponds to $\Lambda_{\rm grav} \sim 4 \pi \left( M_{Pl}^2 m_\gamma / g \right)^{1/3}$, at the cost of bringing the quantum gravity cutoff down to the same scale.
On the other hand, in the regime $g \gtrsim g_*$, the upper bound in Eq.(\ref{eq:Nbound}) is dominant, in turn implying $\Lambda_* \lesssim 64 \pi^3 m_\gamma / g^3$, which again coincides with the upper bound on the cutoff of the anomalous EFT for this range of couplings. Overall, $\Lambda_*$ can be taken to saturate ${\rm min} \left\{ \Lambda_{\rm grav}, \Lambda_{U(1)^3}\right\}$ for all values of $g$. 

\begin{figure}
\centering
	\includegraphics[scale=1.2]{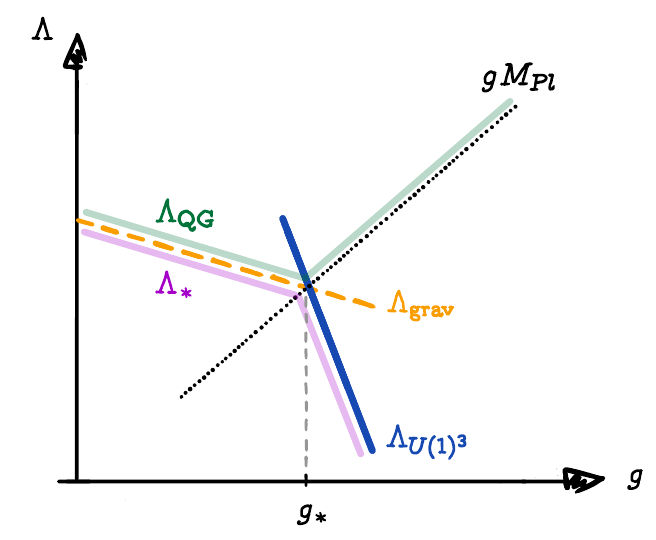}
 	\caption{Upper bound on $\Lambda_*$ (purple line; to be identified with the scale of the radial mode), and corresponding behavior of the quantum gravity scale (green line) when the bound is saturated. For $g \lesssim g_*$, the maximum value of $\Lambda_*$ simultaneously coincides with both $\Lambda_{\rm grav}$ and $\Lambda_{\rm QG}$. For $g \gtrsim g_*$, the upper bound on $\Lambda_*$ coincides with $\Lambda_{U(1)^3}$, while $\Lambda_{\rm QG}$ matches the apparent WGC scale $g M_{Pl}$ in the same limit. (Both axes are in a log scale.)}
\label{fig:both}
\end{figure}

Intriguingly, in the regime where $g \gtrsim g_*$ the scale of quantum gravity $\Lambda_{\rm QG}$ is also subject to a surprising constraint of its own. On the one hand, the upper bound on $N$ in Eq.(\ref{eq:Nbound}) implies $\Lambda_{\rm QG} \gtrsim g M_{Pl}$. On the other hand, after identifying $N \sim \left( 4 \pi M_{Pl} / \Lambda_{\rm QG} \right)^2$, Eq.(\ref{eq:Lambda*N}) can be rewritten as an upper bound on $\Lambda_{\rm QG}$. In combination, these bounds give
\begin{equation} \label{eq:wgc}
	g M_{Pl} \lesssim \Lambda_{\rm QG} \lesssim g {M}_{Pl} \sqrt{ \frac{64 \pi^3 m_\gamma / g^3}{\Lambda_*} } \ .
\end{equation}
Thus, saturating the upper bound on the scale of UV-completion entails bringing the quantum gravity cutoff down to coincide with the scale $g M_{Pl}$ --- the WGC scale as seen in the infrared. Moreover, from Eq.(\ref{eq:Nbound}), $g M_{Pl} \sim g_0^{1/3} M_{Pl}$ in this limit, which is the form of the `magnetic' WGC scale that has been advocated for in \cite{Heidenreich:2016aqi,Heidenreich:2017sim}. The upper bound on $\Lambda_*$, and the behavior of $\Lambda_{\rm QG}$ in the various coupling regimes, are illustrated in Figure~\ref{fig:both}.

There are a variety of noteworthy features in this result. The appearance of the WGC scale in association with the scale of quantum gravity was due not to any direct applications of the WGC, but rather a direct consequence of the large number of species required to probe effective non-compactness of the anomalous $U(1)$.  Even more surprising is the fact that the WGC scale emerges in a largely field-theoretical example featuring a {\it massive} photon, where the direct applicability of the WGC remains conjectural \cite{Reece:2018zvv}. 
Finally, although the WGC scale appearing in Eq.(\ref{eq:wgc}) plays the role of the quantum gravity cutoff only for $g \gtrsim g_*$, note that $g_* \rightarrow 0$ in the limit of a vanishing photon mass.

%%%%%%%%%%%%%%%%%%%%
%%%%%%%%%%%%%%%%%%%%

%%%%%%%%%%%%%%%%%%%%%%%%%%%%%%%%%%%%%%%%%%%%%%%%%%
%%%%%%%%%%%%%%%%%%%%%%%%%%%%%%%%%%%%%%%%%%%%%%%%%%
\section{Conclusions}
\label{sec:discussion}

The study of chiral gauge theories --- the Standard Model being a prime example --- has provided deep insight into general aspects of quantum field theory, as well as concrete understanding of phenomena realized in nature. Moreover, it is in this context that the fundamental differences between Abelian and non-Abelian theories become most apparent. In particular, the existence of mixed $U(1)$-gravitational anomalies --- absent in the non-Abelian case --- provides hope that further scrutiny of this class of theories may provide some insight into the properties of Abelian gauge theories in the context of a gravitational UV-completion.

In this paper, building on the seminal work of \cite{Preskill:1990fr}, we have focused on four-dimensional, massive Abelian gauge theories that are anomaly-free but for which anomaly-cancellation occurs due to fermions appearing at different scales. We have shown that the presence of massive chiral fermions leads to an upper bound on the scale below which a radial mode must become part of the spectrum, and cannot be decoupled. In this class of massive $U(1)$ gauge theories, a St\"uckelberg limit is not allowed.

Maximizing the separation of scales between the massive photon and the scale of the radial mode requires wandering into the (morally) non-compact limit, by introducing a parametrically large ratio of charges in the UV-completion. We have shown that when a $U(1)$-gravitational anomaly is present in the low energy theory, such cavalier behavior is automatically penalized by the quantum gravity scale appearing parametrically below $M_{Pl}$. In turn, this precludes the possibility of falling into the Swampland of theories that remain anomalous above the cutoff scale of the anomalous low energy theory, providing field-theoretic evidence for the existence of a Swampland that is disconnected from the Landscape of consistent EFTs. When the low energy theory also contains a $U(1)^3$ anomaly, there exists a critical value of the gauge coupling, $g_*$. When $g \gtrsim g_*$, saturating the upper bound on the mass of the radial mode comes at the cost of lowering the quantum gravity scale down to $\Lambda_{\rm QG} \sim g M_{Pl}$, which coincides with the WGC scale as seen from the low energy EFT. Our work therefore provides a four-dimensional, field-theoretic example of the WGC scale emerging in the role of a quantum gravity cutoff, in a \emph{massive} Abelian gauge theory, tied to the presence of a large number of species.

Our results resonate with those of \cite{Reece:2018zvv}, and provide qualitative evidence in favor of some of the suggestions contained therein, especially as pertains to the non-decoupling behavior of additional degrees of freedom, as well as to the lowering of the quantum gravity scale in theories with parametrically light vectors. Our work provides strong motivation to further investigate the suggestions of \cite{Reece:2018zvv}, in the context of string theory constructions with massive photons that feature anomalous infrared fermion content, which are also commonplace in the context of string constructions \cite{Dine:1987xk,Aldazabal:1998mr,Ibanez:1998qp,Ibanez:1999pw,Antoniadis:2002cs}.

Finally, our work opens several avenues for further exploration. For example, in dimensions higher than four, there exist additional possibilities for cancelling gauge anomalies that do not require the presence of heavy fermions, such as the ten-dimensional Green-Schwarz mechanism \cite{Green:1984sg} and variations thereof \cite{Witten:1984dg,Dine:1987xk}. Extending our work to study UV-completions in dimensions higher than four could provide more insight into the structure of this class of massive $U(1)$s. Last but not least, we would be remiss to not mention the ubiquity of massive photons in various extensions of the Standard Model, most notably as mediators for dark matter, or as dark matter itself \cite{Nelson:2011sf,Graham:2015rva}. If these theories were to contain an anomaly canceled by massive chiral fermions (see e.g.~\cite{Dror:2017nsg} for work along these lines), our result would apply and could restrict the validity of some of these approaches.

\section*{Acknowledgments}
We thank Arkady Vainshtein and Ken Van Tilburg for helpful conversations, and especially Patrick Draper for collaboration at early stages of this work.
GDK is grateful to KITP and UCSB for hospitality where part of this work was completed.
NC was supported by the U.S.~Department of Energy under the Early Career Award DE-SC0014129 and the Cottrell Scholar Program through the Research Corporation for Science Advancement. 
The research of IGG is funded by the Gordon and Betty Moore Foundation through Grant GBMF7392. 
GDK was supported in part by the 
U.S.~Department of Energy under Grant Number DE-SC0011640.
Research at KITP is supported in part by the National Science Foundation under Grant No.~NSF PHY-1748958.

%%%%%%%%%%%%%%%%%%%%%%%%%%%%%%%%%%%%%%%%%%%%%%%%%%

\bibliography{gaugeanomalies_paper}
	
\end{document}